\documentclass[conference]{IEEEtran}
\IEEEoverridecommandlockouts
\usepackage{cite}

\usepackage{graphicx}%
\usepackage{multirow}%
\usepackage{amsmath,amssymb,amsfonts}%
\usepackage{amsthm}%
\usepackage{mathrsfs}%
\usepackage{xcolor}%
\usepackage[colorlinks=true,
            linkcolor=red,
            urlcolor=blue,
            citecolor=green!50!black!80!]{hyperref}
\usepackage{textcomp}%
\usepackage{manyfoot}%
\usepackage{booktabs}%
\usepackage{algorithm}%
\usepackage{algorithmicx}%
\usepackage{algpseudocode}%
\usepackage{listings}%
\usepackage{enumitem}
\usepackage{url}
\usepackage{color, colortbl}

\usepackage{amsmath,amsfonts,bm}

\def\eqref#1{equation~\ref{#1}}

\def\1{\bm{1}}

\def\vh{{\bm{h}}}

\def\vx{{\bm{x}}}
\def\vy{{\bm{y}}}

\def\mH{{\bm{H}}}

\def\mW{{\bm{W}}}
\def\mX{{\bm{X}}}
\def\mY{{\bm{Y}}}

\DeclareMathAlphabet{\mathsfit}{\encodingdefault}{\sfdefault}{m}{sl}
\SetMathAlphabet{\mathsfit}{bold}{\encodingdefault}{\sfdefault}{bx}{n}

\def\gE{{\mathcal{E}}}

\def\gG{{\mathcal{G}}}

\def\gP{{\mathcal{P}}}

\def\gV{{\mathcal{V}}}

\newcommand{\R}{\mathbb{R}}

\newcommand{\softmax}{\mathrm{softmax}}

\newcommand{\eg}{\textit{e}.\textit{g}., }
\newcommand{\ie}{\textit{i}.\textit{e}., }

\newcommand{\sol}{\textsc{ProtSolM}}
\newcommand{\data}{\textsc{PDBSol}}

\newcounter{bxincomm}
\definecolor{aqua}{rgb}{0.00,0.67,0.80}

\newcounter{tycomm}
\definecolor{orange}{rgb}{0.2,0.7,0.2}

\newcounter{zjcomm}
\definecolor{mypurple}{rgb}{0.6,0,0.6}

\newcounter{todocomm}

\def\BibTeX{{\rm B\kern-.05em{\sc i\kern-.025em b}\kern-.08em
    T\kern-.1667em\lower.7ex\hbox{E}\kern-.125emX}}
\begin{document}

\title{\sol: Protein Solubility Prediction with Multi-modal Features
\thanks{This work was supported by the National Natural Science Foundation of China (11974239; 62302291), the Innovation Program of Shanghai Municipal Education Commission (2019-01-07-00-02-E00076), Shanghai Jiao Tong University Scientific and Technological Innovation Funds (21X010200843), the Student Innovation Center at Shanghai Jiao Tong University, and Shanghai Artificial Intelligence Laboratory.}
}

\author{
\IEEEauthorblockN{Yang Tan}
\IEEEauthorblockA{
\textit{Shanghai Jiao Tong University}\\
Shanghai, China \\
tyang@mail.ecust.edu.cn}
\and 
\IEEEauthorblockN{Jia Zheng}
\IEEEauthorblockA{
\textit{Shanghai Jiao Tong University}\\
Shanghai, China \\
zhengjia2002@sjtu.edu.cn}
\and
\IEEEauthorblockN{Liang Hong}
\IEEEauthorblockA{
\textit{Shanghai Jiao Tong University}\\
Shanghai, China \\
hong3liang@sjtu.edu.cn}
\and
\IEEEauthorblockN{Bingxin Zhou}
\IEEEauthorblockA{
\textit{Shanghai Jiao Tong University}\\
Shanghai, China \\
bingxin.zhou@sjtu.edu.cn}
}

\maketitle

\begin{abstract}
Understanding protein solubility is essential for their functional applications. Computational methods for predicting protein solubility are crucial for reducing experimental costs and enhancing the efficiency and success rates of protein engineering. Existing methods either construct a supervised learning scheme on small-scale datasets with manually processed physicochemical properties, or blindly apply pre-trained protein language models to extract amino acid interaction information. The scale and quality of available training datasets leave significant room for improvement in terms of accuracy and generalization. To address these research gaps, we propose \sol, a novel deep learning method that combines pre-training and fine-tuning schemes for protein solubility prediction. \sol~integrates information from multiple dimensions, including physicochemical properties, amino acid sequences, and protein backbone structures. Our model is trained using \data, the largest solubility dataset that we have constructed. \data~includes over $60,000$ protein sequences and structures. We provide a comprehensive leaderboard of existing statistical learning and deep learning methods on independent datasets with computational and experimental labels. \sol~achieved state-of-the-art performance across various evaluation metrics, demonstrating its potential to significantly advance the accuracy of protein solubility prediction.
\end{abstract}


\section{Introduction}
Protein solubility is a crucial aspect of scientific research and industrial applications. It plays a pivotal role in determining the absorption and metabolism of antibody drugs \cite{vendruscolo2011protein}, enhancing the yield and production efficiency in enzyme engineering \cite{habibi2014review,chan2010learning}, and understanding protein localization and interaction mechanisms \cite{tokmakov2021protein}. However, the majority of expressed non-transmembrane proteins are either insoluble or tend to precipitate or aggregate \cite{fang2013discrimination}. Even among soluble proteins, many have insufficient solubility, limiting their scope of experimental evaluation approaches and applications. The high proportion of insoluble proteins and the high cost of experiments make it impractical to experimentally validate the solubility of every designed protein. Therefore, there is a need for simulation or computational methods to assess protein solubility before experimental validation. 

Early solubility prediction methods were based on physical simulations that used molecular dynamics to calculate the free energy transfer between the condensed and the solution phases \cite{tjong2008prediction,de2011experimental}. However, these methods were limited in accuracy and had high computational costs, restricting their application to large-scale predictions, such as screening designed drug molecules or engineered mutants. Alternatively, statistical methods, such as support vector machine \cite{PROSO,SOLpro,ccSOL} and gradient boosting \cite{Soluprot}, learn the projection of handcrafted protein attributes and solubility from thousands of labeled data. As such models are only capable of processing tabular data, input data have to be manually extracted based on expert knowledge, such as amino acid (AA) composition, charge, and relative surface area. However, due to limitations in dataset scale and model complexity, these methods suffered from significant deficiencies in prediction accuracy and generalization, resulting in limited practical applicability. 

\begin{figure*}
    \centering
    \includegraphics[width=\textwidth]{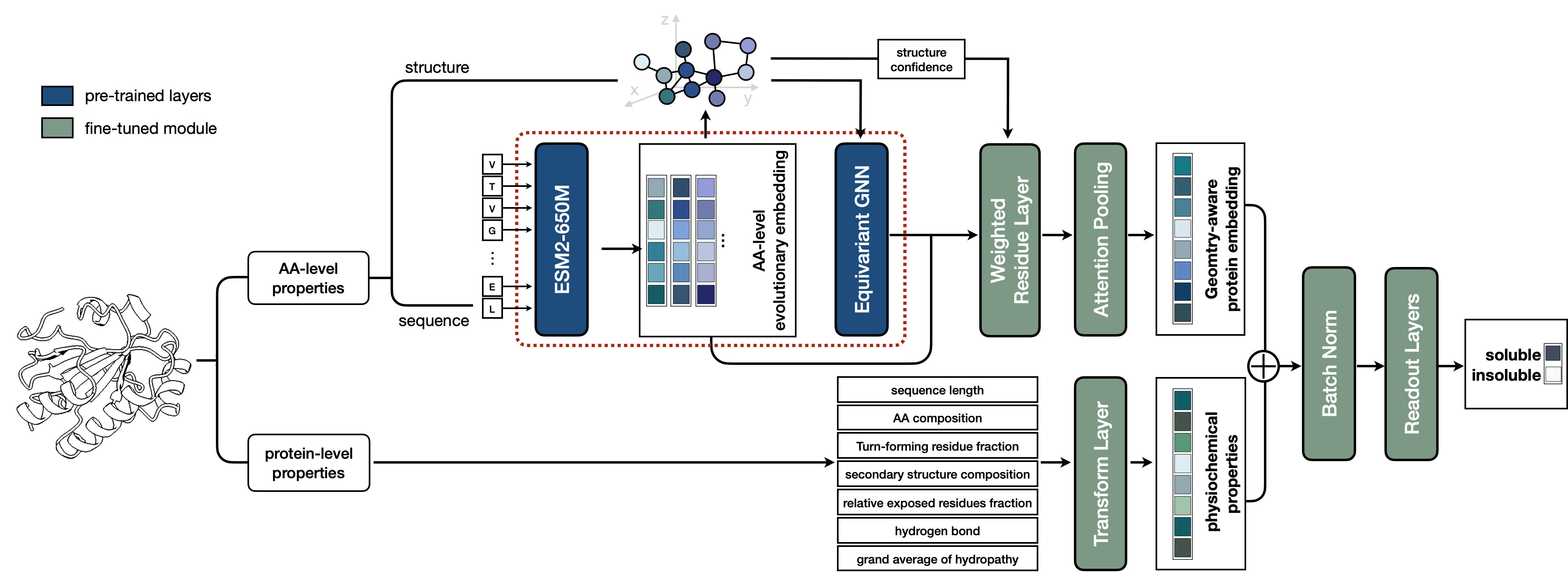}
    \caption{An illustration of \sol~for protein solubility prediction. \sol~employs a pre-training module (in red box) to encode AA-level sequence and structure information, which is then compressed to vector representations with weighted residue layers and attention pooling. At the protein level, hand-crafted physicochemical properties related to solubility are normalized and concatenated to the vector hidden representation. The joint representations are input into fully connected layers for label prediction. These green modules are fine-tuned using \data~to better fit the solubility prediction task.}
    \label{fig:architecture}
\end{figure*}

With the advancement of deep learning technology, many end-to-end prediction methods have been developed. These methods use Transformers \cite{NetSolP,TAPE,ProGAN} or convolutional neural networks \cite{DeepSol,EPSOL,DSResSol} to directly analyze the interactions between AAs. Compared to statistical learning methods, deep neural networks offer higher prediction accuracy and reduce the dependency on feature engineering. However, the limitation of small dataset sizes still persists, causing many models to overfit particular test sets, making their reliability in general applications questionable. Nevertheless, sequence-based encoding methods neglect the local geometric structure of proteins, even though such structures have been found to be significantly associated with protein solubility (\eg $\alpha$-helices \cite{qing2022protein} and surface patches \cite{chan2013soluble}). Although previous studies attempted to encode structure-aware representations by applying graph neural networks, the graphs are constructed through contact maps instead of protein backbone structures \cite{GraphSol,GATSol,DeepMutSol}. This indirect way leads to unnecessary information loss and additional processing steps, potentially reducing the performance of the trained model.

The conflict between the urgent need for protein solubility prediction and the significant deficiencies in current deep learning models as well as training datasets forms a clear research gap. To tackle these issues, this study proposes a new deep learning framework based on pre-training and fine-tuning called \sol, which comprehensively learns and integrates effective vector representations of protein multi-modalities, including sequences, structures, and physicochemical information, for accurate protein solubility prediction. As shown in Fig.~\ref{fig:architecture}, we use the ESM2 framework \cite{esm2} to encode the sequence information of AAs and employ roto-translation equivariant graph neural networks \cite{egnn} to enhance the local interactions based on the protein backbone. This AA-level encoding module, reinforced by geometric information, is pre-trained using a wild-type protein dataset. To further align with the solubility prediction objective, the AA-level encoding is compressed using attention pooling and concatenated with a handful of pre-defined global features that have been proven related to solubility. For a particular protein to approximate the solubility, the joint vector representation is sent to read-out layers for prediction. The second half of the model is fine-tuned on \data, which, to the best of our knowledge, is the largest protein solubility data with comprehensive information on protein entities, including sequence, structure, physicochemical properties, and computational solubility labels. 

In a nutshell, this study makes contributions to the community from three perspectives:
\begin{enumerate}[leftmargin=*]
    \item We propose the first solubility prediction deep learning model that directly integrates protein sequence, structure, and physicochemical properties. Compared to existing prediction methods, the proposed \sol~leverages pre-training and fine-tuning schemes to thoroughly learn the construction logic of natural proteins and explore the relationship between their extracted features and solubility. This enhances the model's overall prediction accuracy and generalization capability (Section~\ref{sec:method}).
    \item We propose the largest, most recent, and most comprehensive solubility training and testing dataset, \data. It includes protein sequences, structures, solubility labels, and solubility-related physicochemical properties. The total number of proteins available for training and analysis exceeds $60,000$ entries (Section~\ref{sec:data}).
    \item We construct a comprehensive leaderboard on the prediction performance of \sol~and existing baseline methods for two independent test datasets with computational and experimental solubility labels. A complete list of statistical learning and deep learning-based methods has been assessed. We also fine-tuned popular pre-trained methods to show their potential for applying to solubility prediction when sufficient training data is available. Notably, our \sol~outperforms existing supervised learning models and fine-tuned self-supervised learning models on various widely-applied evaluation metrics in both test datasets (Section~\ref{sec:results}).
\end{enumerate}

\section{Related Works}
The importance of solubility has given rise to a series of prediction methods. This section reviews three existing approaches, including physics-based calculations, statistical learning, and deep learning methods. The latter two approaches are more suitable for making rapid and accurate predictions on large-scale datasets, but their predictive power depends heavily on the quality of the datasets for training and validation. To this end, the last paragraph reviews the current training and testing datasets available.

\paragraph{Simulation-based Solutions}
One way of predicting protein solubility is through molecular dynamics (MD) simulations. This approach calculates the free energy of transfer from the condensed phase to the solution phase \cite{tjong2008prediction}. However, the difficulty in assessing conformational entropy and solvent effects limits the prediction accuracy of this computationally intensive method. While there are alternative methods to increase simulation accuracy, such as nuclear magnetic resonance experiments \cite{de2011experimental}, the high experimental cost restricts their use to study only a small number of proteins. In modern applications such as protein engineering, it is unrealistic to use MD simulation as a screening tool for predicting the solubility of a large number of protein candidates. 

\paragraph{Statistical Learning Solutions}
The second type of solution optimizes statistical learning-based models from a set of tabular data to find the mapping relationship between the extracted features and solubility. Existing methods often construct discriminative models in machine learning, such as Regularized Linear Regressor \cite{PROSOII,CamSol}, support vector machines \cite{SOLpro,ccSOL,eSOL}, and gradient boosting machines \cite{Soluprot,RPPSP}. The tabular data usually contains manually extracted attributes from protein sequences based on domain knowledge, such as AA composition \cite{SOLpro,Soluprot,RPSP}, fractions of secondary structures \cite{Protein-Sol,SWI,PaRSnIP}, and physicochemical properties of proteins (\eg exposed residue fraction, isoelectric point, and chemical flexibility) \cite{ccSOL,eSOL,RPPSP}. These methods balance computational costs and prediction performance by summarizing and extracting important features from protein data. Additionally, since both the model and input data are relatively simple, it is possible to interpret the input data's strongly related attributes to solubility by analyzing the learned parameters or feature importance of the fitted model, providing a degree of interpretability \cite{ccSOL,SOLart}. Nevertheless, these prediction methods require explicit preprocessing steps, and patterns related to solubility that are excluded in the hand-crafted features significantly limit the model's predictive performance. Moreover, the relatively simple model architecture cannot handle more complex protein data (such as protein structures) and cannot extract more concrete mapping relationships from larger datasets, which further weakens the model's prediction accuracy and generalization.

\paragraph{Deep Learning Solutions}
The development of deep learning methods and the enrichment of protein databases have driven the proposal of more powerful models to directly learn solubility-related patterns from protein sequences and structures. Existing mainstream methods primarily analyze amino acid sequences, using models such as Transformer \cite{NetSolP,TAPE,ProGAN}, LSTM \cite{DeepSoluE,SeqVec}, and CNN \cite{DeepSol,EPSOL,DSResSol} to extract hidden representations of sequences and connect fully connected layers to form a complete supervised learning model for training. Other studies attempted to construct the local interactive relationships between amino acids and use GNN to learn these spatial relationships due to the important role of protein structure in solubility \cite{GATSol,GraphSol,Hybridgnn}. However, these methods rely on contact maps, an intermediate variable from additional processing, which inevitably causes information loss. Moreover, the majority of existing deep learning solutions, whether based on sequences or structures, train a supervised learning model using a limited dataset. This approach not only fails to fully leverage the vast amount of available protein sequences and structural data but also raises questions about whether the model can generalize to accurately predict the solubility of other proteins. 

\paragraph{Training and Evaluation Datasets}
The datasets used to train and validate statistical learning and deep learning models can be divided into two categories: those from solubility experiments and those from other protein databases. The former type typically contains tens to a few thousands of protein instances that are labeled with binarized \cite{NESG} or percentage \cite{eSOL,SC} solubility measurements. Although the solubility labels of these datasets are of higher quality, their relatively small size creates a trade-off between model expressiveness and generalizability. The latter type of datasets, instead, are manually compiled from various sources, such as PDB \cite{PDB} and TargetTrack \cite{TargetTrack}, and can contain up to tens of thousands of data points. The labels for these datasets are usually automatically assigned based on common knowledge. Especially for deep learning models, PROSO II \cite{PROSOII} and its variants \cite{SWI,DeepSol,SKADE} are frequently employed as the training dataset. They perform preprocessing such as redundancy reduction and bias correction on data from various sources. However, these training datasets only include protein entities before 2015 and consist solely of protein sequences and hand-crafted features. Moreover, protein backbone structures are missing in these datasets.

\section{Protein Solubility Prediction with \sol}
\label{sec:method}
Denote $\gP=\{\mX_{\rm AA}, \gG(\gV,\gE), \vx_{\rm prop}\}$ an arbitrary protein of $L$ AAs, which includes the AA sequence $\mX_{\rm AA}\in\R^{L\times 20}$, a graph $\gG(\gV,\gE)$ of the protein backbone structure, and $m$ hand-crafted protein-level physicochemical properties $\vx_{\rm prop}\in\R^{m}$. The task of protein solubility prediction assigns binary labels of soluble and insoluble by
\begin{equation}
    \hat{\vy}=f(\mX_{\rm AA}, \gG(\gV,\gE), \vx_{\rm prop})
\end{equation}
using a trained predictor $f(\cdot)$. This section introduces the formulation of $\gP$ and the proposed \sol~to fit $f(\cdot)$.


%

\subsection{Formulation of Protein $\gG$}
Define $\gG=(\gV,\gE)$ an $k$NN graph of a protein's backbone. Each node $v\in\gV$ represents an AA, and it is connected to up to $k$ nearest nodes within $30$\AA by undirected edges $e\in\gE$. Both nodes and edges are featured with vector attributes, with the node attributes $\mW_V={\rm ESM\_650M}(\mX_{\rm AA})\in\R^{L\times1280}$ are hidden semantic embeddings of AA types extracted by the ESM2 encoder \cite{esm2}. The edge attributes $\mW_E\in\R^{L\times93}$ are defined following \cite{zhou2024protlgn} to feature relationships of connected nodes based on inter-atomic distances, local N-C positions, and sequential position encoding. To preserve the 3D geometry of the protein backbone, $\mX_V\in\R^{L\times3}$ is defined to record 3D coordinates of AAs in the Euclidean space.

The protein-level features $\vx_{\rm prop}\in\R^{42}$ are composed of seven sets of solubility-related features (Table~\ref{tab:features}). The first three sets can be directly extracted from the fasta file, including the fraction of five charged AAs (cysteine, aspartate, glutamic acid, arginine, and histidine) relative to the total number of amino acids, the fraction of turn-forming AAs (aspartic acid, glycine, and proline) relative to the total number of amino acids, and the grand average of hydropathy (GRAVY) index, which reports the average of the hydropathy values of all the amino acids in a protein sequence \cite{kyte1982simple}. The fractions of secondary structures are calculated from DSSP classifications \cite{DSSP} with 3 states and 9 states (including an undefined state). The fraction of exposed residues \cite{PaRSnIP} lists the fractions of AAs with relative solvent accessibility (RSA) cutoffs between $5\%$ and $100\%$. The amount and density of hydrogen bonds are predicted by \textsc{MDTraj} \cite{MDTraj}. The structure confidence is represented by average pLDDT from ESMFold \cite{esm2} prediction.

\begin{table}[!t]
\caption{Protein-level physiochemical features.}
\vspace{-4mm}
\label{tab:features}
\begin{center}
\begin{tabular}{lcc}
    \toprule
    \textbf{Feature} & \textbf{Dimension} & \textbf{Source} \\
    \midrule
    fraction of charged AA (C, D, E, R, H) & 5 & fasta\\
    fraction of turn-forming residues (N, G, P) & 1 & fasta\\
    GRAVY index & 1 & fasta\\
    fraction of secondary structure & 12 & DSSP\\
    fraction of exposed residues & 20 & DSSP\\
    hydrogen bonds & 2 & MDTraj\\
    structure confidence & 1 & PDB\\
    \bottomrule
\end{tabular}
\end{center}
\end{table}


\subsection{Model Pipeline}
As illustrated in Fig.~\ref{fig:architecture}, \sol~consists of two modules of AA-level feature encoding and protein-level solubility prediction. The former (in blue boxes) is pre-trained with self-supervised learning, and the latter (in green boxes) is fine-tuned with solubility labels.

\paragraph{Pre-training Module}
The AA-level encoding module is pre-trained to extract adequate structure-aware evolutionary embedding for proteins. The evolutionary embedding $\mW_V$ is encoded from protein sequences by ESM2 \cite{esm2}, which is used as node features for protein graphs. For graph representation learning, EGNN layers \cite{satorras2021n} are employed in consideration of rotation equivariance and translation invariance of protein representation. This module is trained on a denoising task \cite{zhou2024protlgn,protssn}, where the input AA types for ESM2 are perturbed with multinomial noise, \ie 
an AA in the input protein sequence has a chance of $p$ to mutate to one of $20$ AAs (including itself) with the replacement distribution defined by the frequency of AA types observed in wild-type proteins. The hidden embedding of AAs $\mH_G$ is projected by a fully connected layer to a 20-dimensional output $\mY_0$ that indicates the probability of each node being one of the 20 types of AA. The model is trained to minimize the cross-entropy of the predicted and ground-truth AA types.

\paragraph{Fine-tuning Module}
The fine-tuning module utilizes the AA-level hidden representation $\{\mW_V,\mH_G\}$ from the pre-trained module and trains the remaining network layers with the solubility dataset (Section~\ref{sec:data}). The AA-level sequence and structure representation is summarized by a weighted residual connection with the pLDDT penalty term, \ie
\begin{equation}
    \mH_{AA} = \mW_V + {\rm pLDDT} \times \mH_G.
\end{equation}
The combined embedding $\mH_{AA}$ is then processed to obtain protein-level representation by an attention pooling layer \cite{tan2023peta}:
\begin{equation}
    \hspace{-2mm}\vh_{P} = {\rm AttnPool}(\mH_{AA})=\softmax(\mathrm{Conv}(\mH_{AA}))\cdot\mH_{AA},
\end{equation}
where $\mathrm{Conv}(\cdot)$ represents a 1-dimensional convolution along the dimension of the AA sequence and $\cdot$ computes the weighted average of AA embeddings. 

The output vector $\vh_{P}$ is concatenated to $\vh_{\rm prop}$, the transformed protein-level attributes by linear projection $\mW_{prop}\in\R^{42\times512}$ and batch normalization, \ie 
\begin{equation}
\begin{aligned}
    \vh&={\rm concat}(\vh_{P},\vh_{\rm prop}),\\
    \text{where}\;\vh_{\rm prop}&=\mW_{prop} \cdot {\rm BatchNorm}(\vx_{\rm prop}).
\end{aligned}
\end{equation}
The resulting multi-level representation of the protein is sent to fully connected layers to output binary classification
\begin{equation}
    \hat{\vy}=\mW_o({\rm DropOut}({\rm \text{ReLU}}(\vh)),
\end{equation}
where the ReLU activation and the dropout layer are for avoiding overfitting, and the final projection $\mW_o\in\R^{1792\times2}$ is added for predicting the labels. 


\begin{table}[!t]
\caption{Source of raw data for \data.}
\label{tab:sourceData}
\vspace{-4mm}
\begin{center}
\begin{tabular}{lrrr}
    \toprule
    \textbf{Dataset} & \textbf{\# Total} & \textbf{\# Soluble} & \textbf{\# Insoluble}\\
    \midrule
    UniProtKB & 4,337 & 4,337 & 0\\
    TargetTrack & 468,406 & 287,844 & 180,562\\
    PDB & 402,059 & 402,059 & 0\\
    PRSP-2k & 2,001 & 1,000 & 1,001\\
    \midrule
    \data~(raw) & 311,635 & 198,164 & 113,471\\
    \bottomrule
\end{tabular}
\end{center}
\end{table}

\section{Dataset and Material}
\label{sec:data}
As discussed previously, existing datasets for solubility are neither large enough nor up-to-date. We thus prepare \data, a new dataset for training solubility models.

\subsection{Source of Raw Data}
\data~obtains raw data from diverse publicly available protein databases (Table~\ref{tab:sourceData}), including:
\begin{itemize}[leftmargin=*]
    \item \textbf{UniProtKB} \cite{10.1093/nar/gkl929}: a large database of protein sequences with high-quality, hand-annotated protein records. Following \cite{SOLpro}, we selected `\texttt{Reviewed}' proteins labeled with `\texttt{E. coli enzymes}' or `\texttt{S. cerevisiae enzymes}'. A total of $4,337$ `\texttt{soluble}' proteins were obtained by this approach.
    \item \textbf{TargetTrack} \cite{TargetTrack}: a comprehensive database of proteins with the results of structural experiments and corresponding status history. We picked both positive and negative instances based on their experimental status \cite{PROSOII}. All records that reached soluble or subsequent status were considered `\texttt{soluble}'. For `\texttt{insoluble}' samples, we included terminated records that were highly likely to have failed to be expressed or purified. The total number of soluble and insoluble proteins was $287,844$ and $180,562$, respectively. 
    \item \textbf{PDB} \cite{PDB}: a vast collection of experimental protein structures. We selected $402,059$ `\texttt{soluble}' proteins that were encoded into \texttt{plasmids} and expressed in \texttt{E. coli}.
    \item \textbf{PRSP-2k} \cite{chang2014bioinformatics}: a balanced solubility dataset with $2,001$ instances \cite{PROSO,PROSOII}. The source of data and processing step is similar to \data, we thus merge them into our dataset.
\end{itemize}

After removing repetitive proteins, the initial \data~collects $198,164$ soluble and $113,471$ insoluble proteins. 

\definecolor{Gray}{gray}{0.9}
\begin{table}[!t]
\caption{Preprocessing details on \data.}
\label{tab:DataProcess}
\vspace{-4mm}
\begin{center}
\resizebox{\linewidth}{!}{
\begin{tabular}{lrrr}
    \toprule
    \textbf{Dataset} & \textbf{\# Total} & \textbf{\# Soluble} & \textbf{\# Insoluble}\\
    \midrule
    \data~(raw) & 311,635 & 198,164 & 113,471\\
    \;$-$ Non-protein Entities & 302,214 & 189,551 & 112,663\\
    \;$-$ Biased Sequential Components & 280,297 & 169,507 & 110,790\\
    \;$-$ Sequence Redundancy & 70,167 & 37,678 & 32,489\\
    \;$-$ Transmembrane Proteins & 67,984 & 35,495 & 32,489\\
    \;$-$ Biased Length \& Class & 64,598 & 33,763 & 30,835\\
    \midrule
    \data-train & 58,138 & 30,419 & 27,719\\
    \data-valid & 3,230 & 1,669 & 1,561\\
    \data-test & 3,230 & 1,675 & 1,555\\
    \bottomrule
\end{tabular}
}
\end{center}
\end{table}

\subsection{Data Preprocessing}
The bias and redundancy in the raw data need to be removed through additional processing steps before being used by a model for pattern recognition. 

\begin{enumerate}[leftmargin=*]
    \item \textbf{Non-protein Entities}: We remove samples annotated with `\texttt{virus}', `\texttt{DNA}', or `\texttt{RNA}' to exclude non-protein crystal structures from PDB. Sequences consisting solely of A', T', U', C', and G' strings, as well as those with more than $5$ consecutive `X' stings, are also excluded. 
    \item \textbf{Biased Sequential Components}: We remove sequences containing tags such as ‘MGSSHHHH’, ‘MHHHHHHS’, and ‘MRGSHHHH’. These His-tags, used for affinity purification, are highly correlated with soluble proteins \cite{His-tag}.
    \item \textbf{Sequence Redundancy}: To eliminate redundant samples and avoid data leakage, we employed \textsc{MMSeqs2} \cite{steinegger2017mmseqs2} to cluster sequences at a $25\%$ identity cutoff. Sequences with fewer than $20$ AAs or more than $2,000$ AAs were also removed due to their rarity.
    \item \textbf{Transmembrane Proteins}: Given the unique solubility properties of transmembrane proteins, we use \textsc{DeepTMHMM} \cite{hallgren2022deeptmhmm} to exclude predicted transmembrane proteins (labeled as `\texttt{TM}', `\texttt{SP+TM}', or '\texttt{BETA}').
    \item \textbf{Biased Length \& Class}: Protein solubility is generally negatively correlated with its length \cite{albu2015feature}. To remove this bias, we divide proteins into eight groups based on their sequence length and balance the group size.
\end{enumerate}

In total, we obtain $64,598$ processed samples for the complete \data. We randomly split $5\%$ for validation and $5\%$ for test. The resulting training, validation, and test dataset contains $58,138$, $3,230$, and $3,230$ proteins, respectively. The full detail is provided in Table~\ref{tab:DataProcess}. Except for the sequence details and the solubility label, \data~also provides predicted structures for each protein by ESMFold \cite{esm2}.

\subsection{Overview of the Training and Testing Datasets}
For the processed \data, we use \data-train to fine-tune \sol, as well as other zero-shot baseline models (see Section~\ref{sec:results} for more details). The validation set is used for model selection, and the independent \data-test constructs a standard test set for evaluating the performance of different baseline models. For a more comprehensive comparison, we collect external test datasets from three open benchmarks used in the literature, including \textbf{ESOL}-agg \cite{eSOL}, \textbf{NESG}-SoluProt \cite{Soluprot}, and \textbf{DESG}-DEResSol \cite{DSResSol,NetSolP}. The labels in the latter two datasets are binarized. For ESOL, we follow \cite{niwa2009bimodal} and define scores  lower than $30$ as insoluble samples and higher than $70$ as soluble samples. We also compare the sequence identity of proteins in the three external datasets and remove $75$ repetitive samples. Note that a $25\%$ sequence identity upper limit between \data-train and the external test dataset is guaranteed to avoid data leakage. Any similar sequences from the training set were removed during preprocessing.


\begin{table}[!t]
\caption{Source of External Test Dataset.}
\label{tab:EvaluationData}
\vspace{-4mm}
\begin{center}
\begin{tabular}{lrrr}
    \toprule
    \textbf{Dataset} & \textbf{\# Total} & \textbf{\# Soluble} & \textbf{\# Insoluble}\\
    \midrule
     ESOL-agg & 2,155 & 951 & 1,204\\
     NESG-SoluProt & 1,784 & 1,052 & 732\\
     NESG-DSResSol & 3,640 & 1,817 & 1,823\\
    \midrule
    External Test Dataset & 7,579 & 3,820 & 3,759\\
    \bottomrule
\end{tabular}
\end{center}
\end{table}

\begin{table*}[!t]
\caption{Performance comparison on the standard test dataset.}
\label{tab:testResult}
\vspace{-4mm}
\begin{center}
\begin{tabular}{lcrcccccccccc}
    \toprule
    \multirow{2}{*}{\textbf{Model}} & \multirow{2}{*}{\textbf{Version}} & \multicolumn{5}{c}{\textbf{Standard Test Dataset} (\data-test)} & \multicolumn{5}{c}{\textbf{External Test Dataset}} \\ \cmidrule(lr){3-7} \cmidrule(lr){8-12}
    &  & ACC & Precision & Recall & AUC & MCC & ACC & Precision & Recall & AUC & MCC \\
    \midrule
    \rowcolor{Gray}
    \multicolumn{12}{c}{Supervised Domain Models}\\
    \midrule
    DeepSoluE \cite{DeepSoluE} & - & 0.578 & 0.595 & 0.588 & 0.610 & 0.156 & 0.603 & 0.623 & 0.539 & \textbf{0.636} & 0.208 \\
    ccSOL omics \cite{ccSOL} & - & 0.533 & 0.545 & 0.599 & 0.610 & 0.061 & 0.522 & 0.524 & 0.578 & 0.530 & 0.044 \\
    SoluProt \cite{Soluprot} & - & 0.646 & 0.634 & 0.750 & 0.733 & 0.292 & \textcolor{red}{\textbf{0.613}} & 0.605 & \textcolor{violet}{\textbf{0.669}} & \textcolor{violet}{\textbf{0.655}} & \textcolor{violet}{\textbf{0.227}} \\
    SKADE \cite{SKADE} & - & 0.656 & 0.773 & 0.476 & 0.727 & 0.349 & 0.503 & 0.530 & 0.118 & 0.568 & 0.018 \\
    Camsol \cite{CamSol} & - & 0.580 & 0.605 & 0.546 & - & 0.163 & 0.594 & 0.626 & 0.485 & - & 0.195 \\
    NetSolP \cite{NetSolP} & - & 0.540 & 0.539 & \textbf{0.785} & 0.572 & 0.071 & 0.544 & 0.527 & \textcolor{red}{\textbf{0.943}} & \textcolor{red}{\textbf{0.695}} & 0.137 \\
    DSResSOL \cite{DSResSol} & - & 0.658 & 0.732 & 0.538 & 0.717 & 0.335 & 0.530 & 0.600 & 0.201 & 0.603 & 0.086 \\
    \midrule
    \rowcolor{Gray}
    \multicolumn{12}{c}{Fine-tuned Protein Language Models}\\
    \midrule
    \multirow{2}{*}{ESM2 \cite{esm2}} & t30\_150M & 0.646 & 0.657 & 0.650 & 0.646 & 0.292 & 0.598 & 0.612 & 0.554 & 0.600 & 0.198\\
    & t33\_650M & 0.648 & 0.670 & 0.618 & 0.649 & 0.298 & 0.597 & 0.620 & 0.514 & 0.597 & 0.197 \\
    [2mm]
    \multirow{2}{*}{ProtBert \cite{elnaggar2021prottrans}} & uniref & 0.623 & 0.632 & 0.670 & 0.621 & 0.244 & 0.568 & 0.568 & \textbf{0.593} & 0.568 & 0.135\\
    & bfd & 0.642 & 0.662 & 0.617 & 0.642 & 0.285 & 0.592 & 0.613 & 0.520 & 0.593 & 0.188\\
    [2mm]
    \multirow{2}{*}{ProtT5 \cite{elnaggar2021prottrans}} & xl\_uniref50 & 0.647 & 0.671 & 0.611 & 0.645 & 0.269 & 0.570 & 0.598 & 0.451 & 0.571 & 0.147\\
    & xl\_bfd & 0.632 & 0.664 & 0.573 & 0.634 & 0.269  & 0.582 & 0.605 & 0.491 & 0.582 & 0.168\\ 
    [2mm]
    \multirow{2}{*}{Ankh \cite{elnaggar2023ankh}} & base & 0.636 & 0.657 & 0.605 & 0.636 & 0.273  & 0.593 & 0.616 & 0.512 & 0.594 & 0.191\\
    & large & 0.648 & 0.657 & 0.655 & 0.648 & 0.295  & 0.602 & 0.614 & 0.568 & 0.602 & 0.205\\
    \midrule
    \rowcolor{Gray}
    \multicolumn{12}{c}{Fine-tuned Multi-modal and Multi-level Model (Ours)}\\
    \midrule
    \multirow{3}{*}{\sol} & k10\_h512 & \textbf{0.789} & \textbf{0.782} & \textcolor{red}{\textbf{0.824}} & \textcolor{violet}{\textbf{0.790}} & \textcolor{violet}{\textbf{0.578}} & \textbf{0.606} & \textcolor{violet}{\textbf{0.655}} & 0.462 & 0.607 & \textbf{0.224}\\
    & k20\_h512 & \textcolor{violet}{\textbf{0.794}} & \textcolor{red}{\textbf{0.804}} & \textcolor{violet}{\textbf{0.798}} & \textcolor{red}{\textbf{0.794}} & \textcolor{red}{\textbf{0.588}} & \textcolor{violet}{\textbf{0.607}} & \textcolor{red}{\textbf{0.664}} & 0.446 & 0.608 & \textcolor{red}{\textbf{0.229}}\\
    & k30\_h512 & \textcolor{red}{\textbf{0.796}} & \textcolor{red}{\textbf{0.804}} & 0.783 & \textbf{0.789} & \textbf{0.577} & 0.601 & \textbf{0.652} & 0.446 & 0.602 & 0.215\\
    \bottomrule\\[-2.5mm]
    \multicolumn{7}{l}{$\dagger$ The top three are highlighted by \textbf{\textcolor{red}{First}}, \textbf{\textcolor{violet}{Second}}, \textbf{Third}.}
    \end{tabular}
\end{center}
\end{table*}

\section{Experiment}
\label{sec:results}
We conduct comprehensive testing and comparison of a series of open-source statistical learning and deep learning models on the standard test dataset and external test datasets. All experiments and protein folding with ESMFold were conducted on 8 80GB-VRAM A800 GPUs. The implementation can be found at \url{https://github.com/tyang816/ProtSolM}.

\subsection{Experimental Protocol}
\paragraph{Training Setup}
\sol~is constructed with the modules introduced in Section~\ref{sec:method}. The pre-trained module employs \texttt{ProtSSN-k20\_h512} \cite{protssn} for joint structure and sequence embedding, with the evolutionary embedding extracted by \texttt{ESM-650M} \cite{esm2}. This step outputs an AA-level hidden representation of $512$ dimension. For the fine-tuning module, we used the AdamW optimizer with a learning rate set at $0.0005$, a weight decay of $0.01$, and a dropout rate of $0.1$ for the output layer. To ensure stable training costs and avoid memory explosion, we adopted a dynamic batching approach, filling each batch up to $16,000$ tokens to ensure $n \times l \leq 16,000$, with $n$ being the number of sequences and $l$ being the maximum length of sequences at the current batch. For the vanilla PLMs, each batch contains a max of $80,000$ tokens, including the padding tokens. The maximum training epoch was set to $30$ with a patience of $5$. The monitor for early stopping for all experiments is based on the accuracy (ACC) of the validation dataset.

\paragraph{Baseline Methods}
The solubility prediction performance of \sol~is compared on two types of models. The first is supervised  machine learning or deep learning models, including DeepSoluE \cite{DeepSoluE}, ccSOL omics\cite{ccSOL}, SoluProt \cite{Soluprot}, SKADE \cite{SKADE}, Camsol \cite{CamSol}, NetSolP \cite{NetSolP}, and DSResSOL \cite{DSResSol}. The second category contains fine-tuned protein language models, including different versions of ESM2 \cite{esm2}, ProtBert \cite{elnaggar2021prottrans}, and encoder-decoder architecture models: ProtT5 \cite{elnaggar2021prottrans}, Ankh \cite{elnaggar2023ankh}. For the second type methods, we fine-tune them with \data~on the published checkpoints from self-supervised learning procedures. All models are implemented based on the officially released program or web server listed in Appendix~\ref{app:source}.


\subsection{Baseline Comparison}
\begin{figure}[!t]
    \centering
    \includegraphics[width=0.49\linewidth]{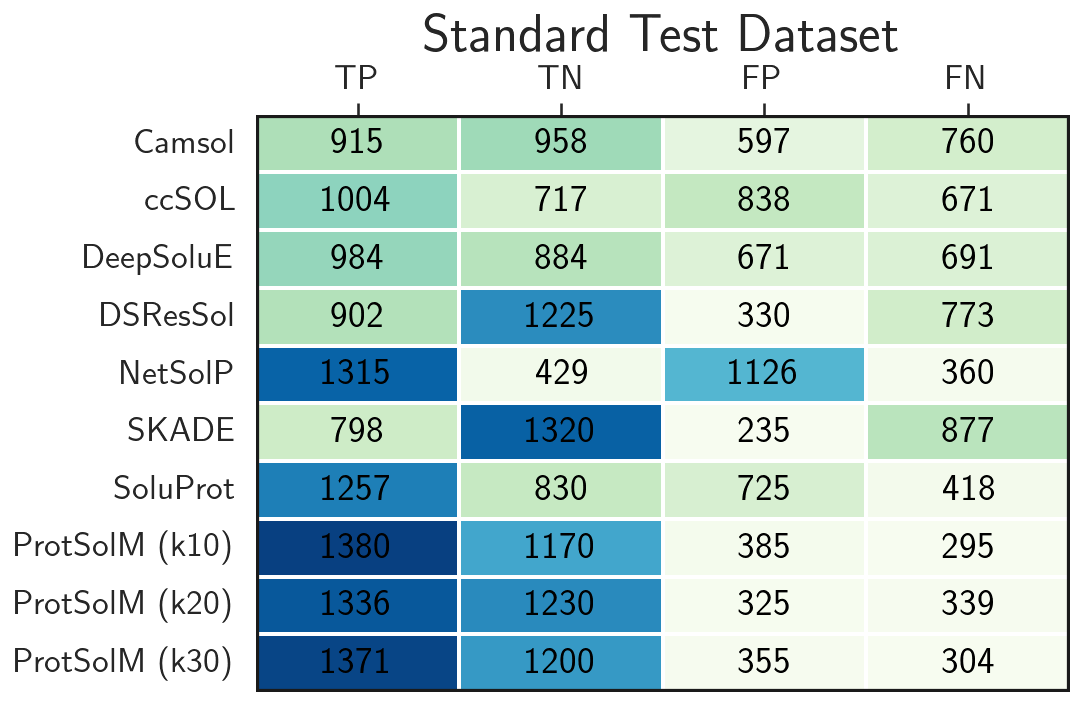}\hfill
    \includegraphics[width=0.49\linewidth]{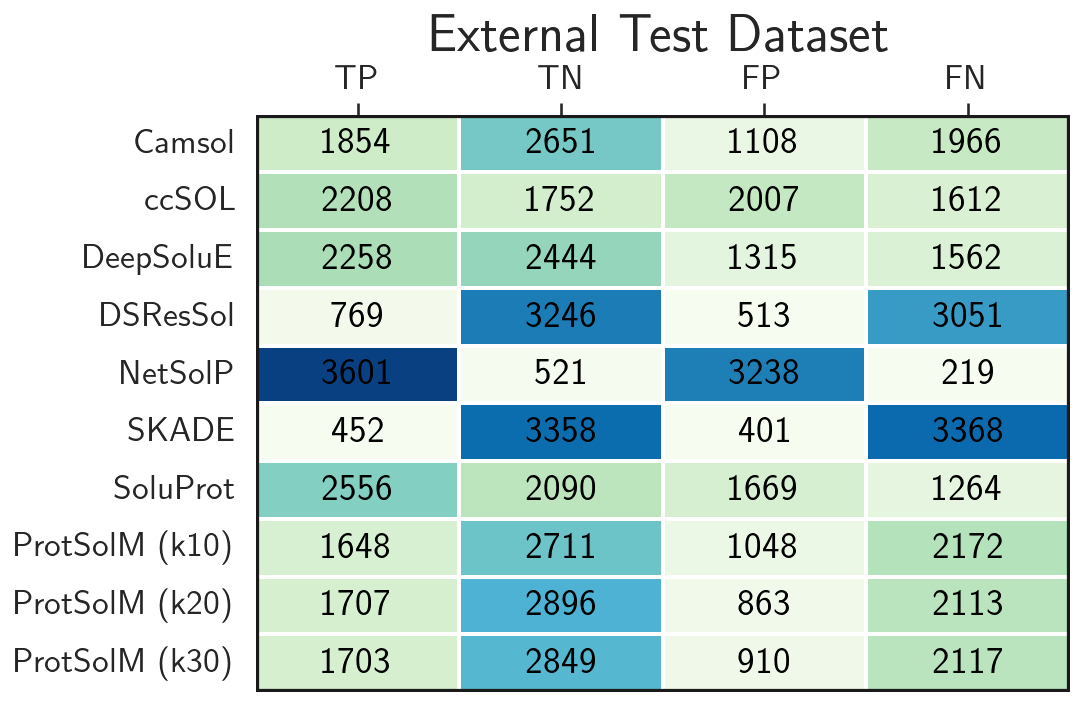}
    \caption{Confusion matrix of domain models (up) and \sol~(down) on standard test set (left) and external test set (right).}
    \label{fig:confusion_heatmap}
\end{figure}

Table~\ref{tab:testResult} reports the overall performance of baseline methods on the standard test dataset and external test datasets. Various evaluation metrics are used to provide a comprehensive assessment of the prediction performance, including accuracy (ACC), precision, recall, the area under the ROC curve (AUC), and Matthew's correlation coefficient (MCC). Unless specified by the authors, all predictions are binarized with $0.5$ classification threshold. Different versions of \sol~achieved the overall best performance across the five evaluation metrics on both test datasets. Notably, the prediction performance of \sol~significantly surpassed other models on \data-test. Moreover, fine-tuned protein language models generally outperformed supervised domain models, indicating that increasing the model size and training dataset (even with unlabeled pre-training data) helps improve solubility prediction performance. On the other hand, most models had score differences within $0.2$ on the external test datasets, with \sol~outperforms baseline methods on ACC, precision, and MCC. The high scores achieved by domain models on recall and AUC, aside from the possibility that some domain models might have been picked for a specific test dataset, are due to the highly biased results predicted by these models. For instance, ccSOL and DSResSol tend to provide more negative predictions, and NetSolP reports more positive predictions than others (Fig.~\ref{fig:confusion_heatmap}). 

\subsection{Additional Investigation}
\begin{figure}[!t]
    \centering
    \includegraphics[width=0.48\linewidth]{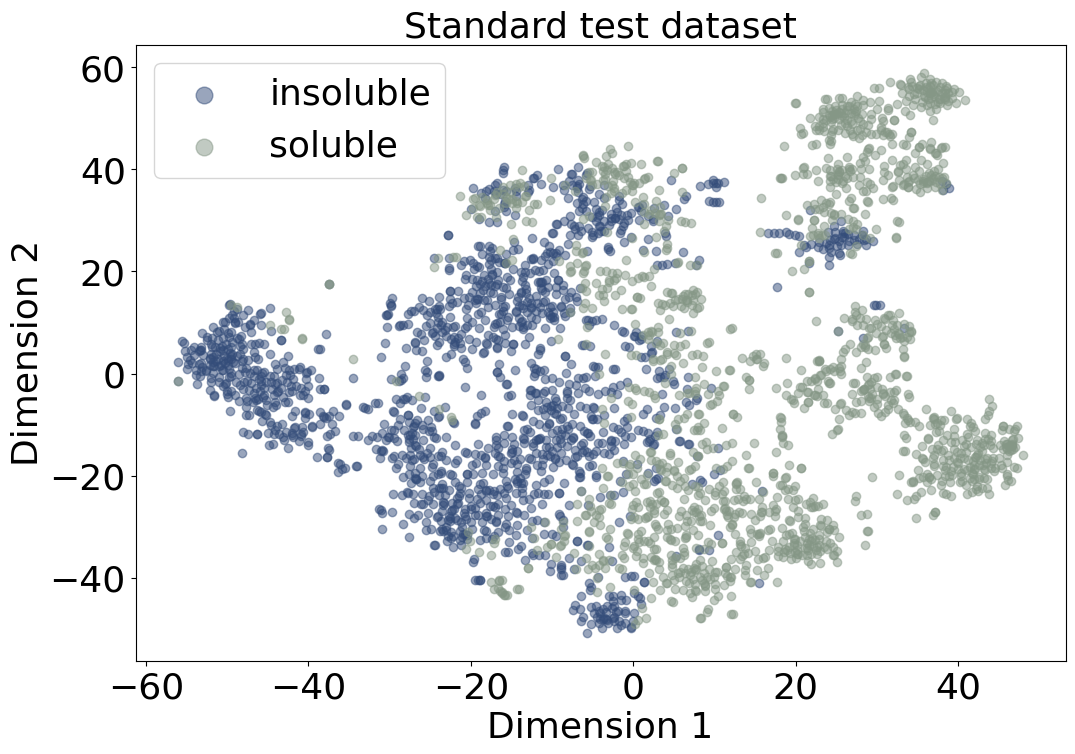}\hfill
    \includegraphics[width=0.48\linewidth]{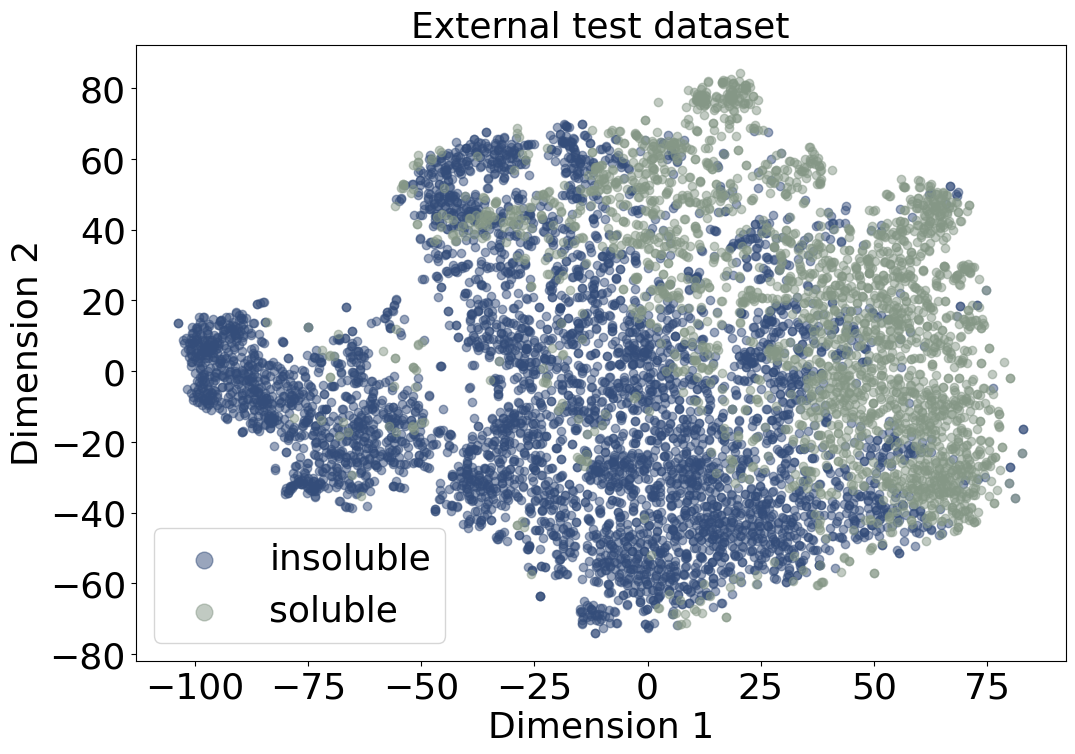}
    \caption{t-SNE visualization of \sol's protein-level representation for standard test dataset (left) and external test dataset (right).}
    \label{fig:t-sne}
\end{figure}

Table~\ref{tab:ablation} reports the performance \data-test by the ablation models of \sol. We validate the combinatorial effect of the three main components in the fine-tuned module, including the pLDDT penalty (PP) used in the weighted residual layer, the hand-crafted protein-level physicochemical features (feature), and the attention pooling layer (AttnPool). The prediction performance drops most significantly when substituting the attention pooling with an average pooling layer. Removing PP and physicochemical features, although less significant, also results in worse performance in the overall prediction. This observation implies that the hidden sequence and structure representation extracted by large pre-trained models provide abundant information for understanding proteins. However, the prediction model on a specific downstream task can still benefit from adding high-quality supplementary features. Moreover, we investigate the expressivity of learned protein embedding by \sol. The results are visualized through t-SNE in Fig.~\ref{fig:t-sne}. On both test datasets, a dot represents a protein-level hidden representation (before the final readout layer) encoded by \sol. Overall, the soluble and insoluble samples are separable for both datasets.

\begin{table}[!t]
\caption{Ablation study of \sol~on \data-test.}
\centering
\resizebox{\linewidth}{!}{%
    \begin{tabular}{lrrrrr}
    \toprule
    Component & ACC & Precision & Recall & AUC & MCC \\ 
    \midrule
    \rowcolor{Gray}
    full \sol & \textbf{0.794} & \textbf{0.804} & 0.798 & \textbf{0.794} & \textbf{0.588} \\
    w/o PP & 0.791 & 0.792 & 0.811 & 0.791 & 0.582 \\
    w/o AttnPool & 0.779 & 0.773 & 0.811 & 0.777 & 0.557 \\
    w/o feature & 0.791 & 0.791 & 0.813 & 0.791 & 0.582 \\
    w/o feature+PP & 0.787 & 0.776 & \textbf{0.828} & 0.785 & 0.574 \\
    w/o feature+PP+AttnPool & 0.773 & 0.763 & 0.815 & 0.771 & 0.545 \\ \bottomrule
    \end{tabular}%
 }
\label{tab:ablation}
\end{table}

\section{Discussion and Conclusion}
\label{sec:discussion}
In this study, we addressed the research gap for accurate and generalized protein solubility prediction by proposing~\sol, a novel deep learning framework that integrates multi-modalities of proteins including sequence, structure, and physicochemical properties. \sol~is pre-trained on large protein sequence and structure datasets, and fine-tuned with \data~to fit the solubility prediction task, which is the largest and most comprehensive dataset to date, which includes over 60,000 protein sequences, structures, and solubility labels. Our benchmark tests against existing SOTA machine learning and deep learning methods on multiple experimentally-based open benchmarks across various evaluation metrics. 

Despite advancements, several challenges are left to address in the future. First, improvements on the interpretability of \sol~is needed. While deep learning models in general lack explainability, developing tools to understand the specific contribution of different features to solubility predictions could enhanced the usefulness and reliability in practical applications. Also, the current version of \sol~is trained on computational labels to maximize the use of available data. It is possible to refine the model with a few high-quality labeled data, such as experimental results. 

In summary, \sol~represents a significant step forward in protein solubility prediction, combining advanced deep learning techniques with comprehensive dataset utilization. Our model not only addresses existing limitations but also opens new avenues for research and application in protein engineering, drug development, and biotechnology. By providing a more accurate and generalized solubility prediction tool, \sol~has the potential to accelerate scientific discovery and innovation in these fields



\bibliographystyle{IEEEtran}
\bibliography{1reference.bib}

\clearpage
\appendices

\section{Additional Experimental Results}
This section provides more experimental results in supplementary of the main text.

\begin{figure}[h]
    \centering
    \includegraphics[width=\linewidth]{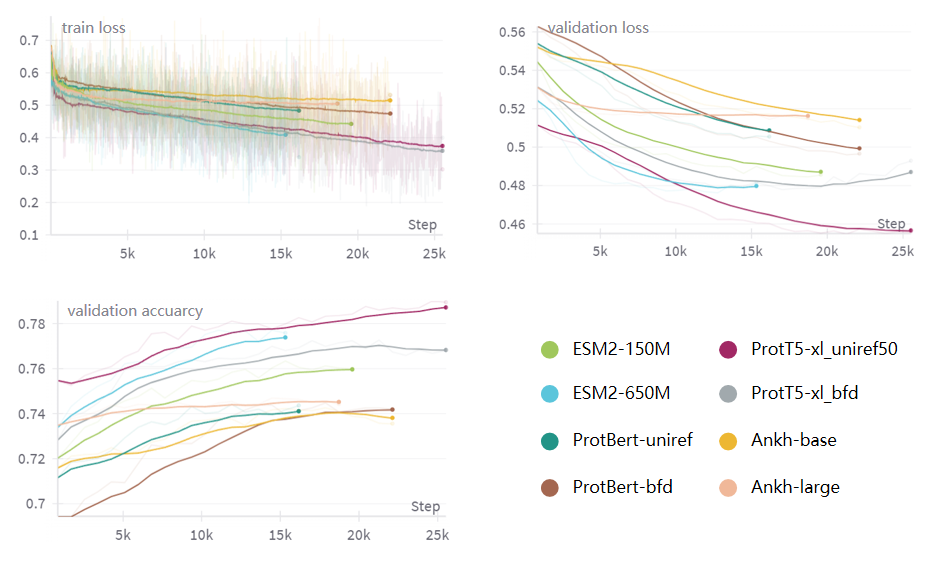}
    \caption{Learning curves of fine-tuning protein language models.}
    \label{fig:plm_curve}
\end{figure}

Fig.~\ref{fig:ablation_curve}-\ref{fig:plm_curve} visualizes the learning curves of ablation models and the fine-tuned models, respectively. From Fig. \ref{fig:ablation_curve}, we can see that attention pooling has a greater impact on the performance of downstream tasks, followed by features and PP. From Fig. \ref{fig:plm_curve}, we can see that ProtT5-xl\_uniref50 performs best on the validation dataset, but it can be seen that there is no significant difference between the performance of ProtT5-xl\_uniref50 and other models on the standard test dataset and the additional test dataset.

\begin{figure}[h]
    \centering
    \includegraphics[width=\linewidth]{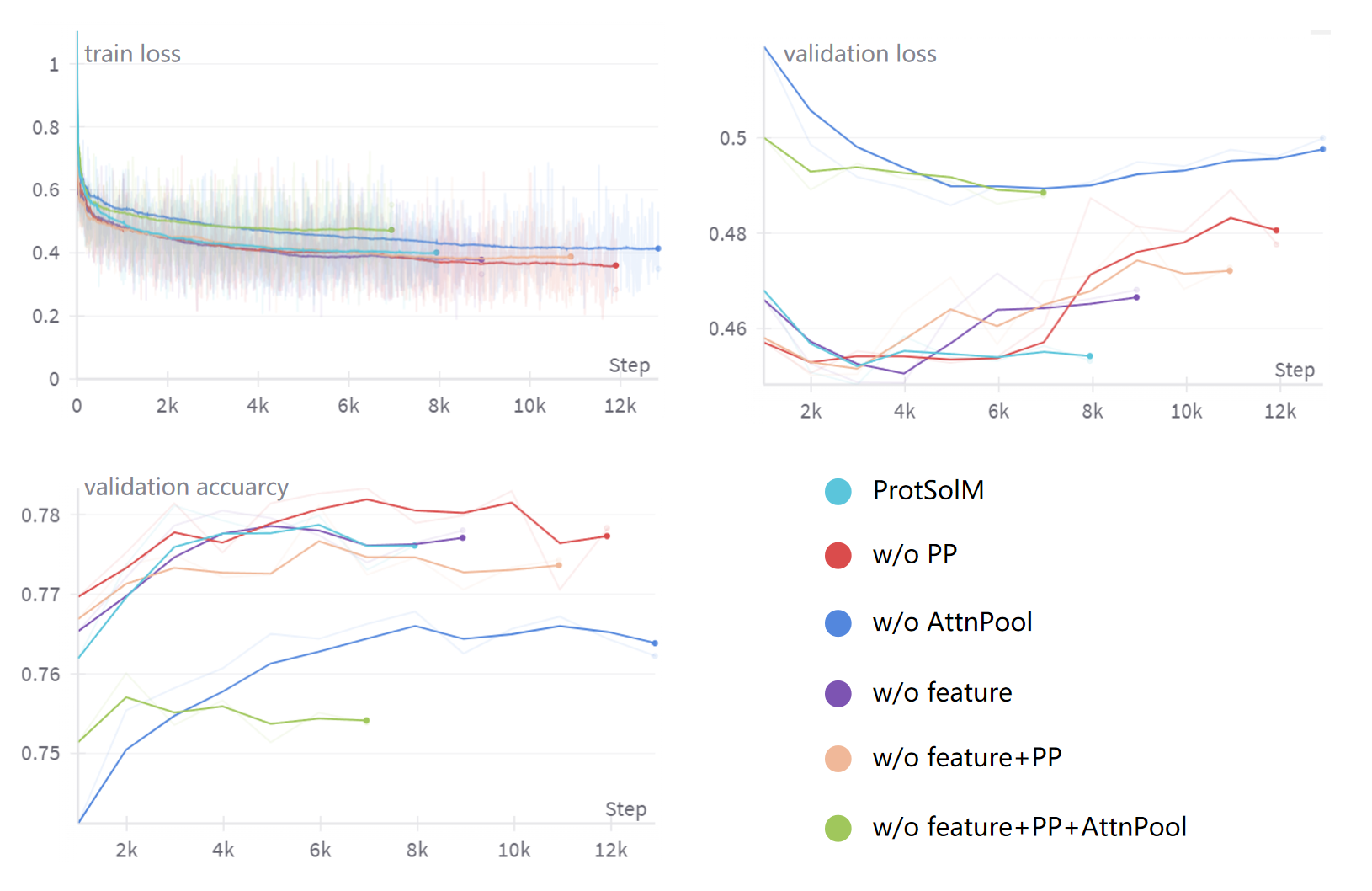}
    \caption{Learning curves of ablation study.}
    \label{fig:ablation_curve}
\end{figure}

\begin{figure}[h]
    \centering
    \includegraphics[width=\linewidth]{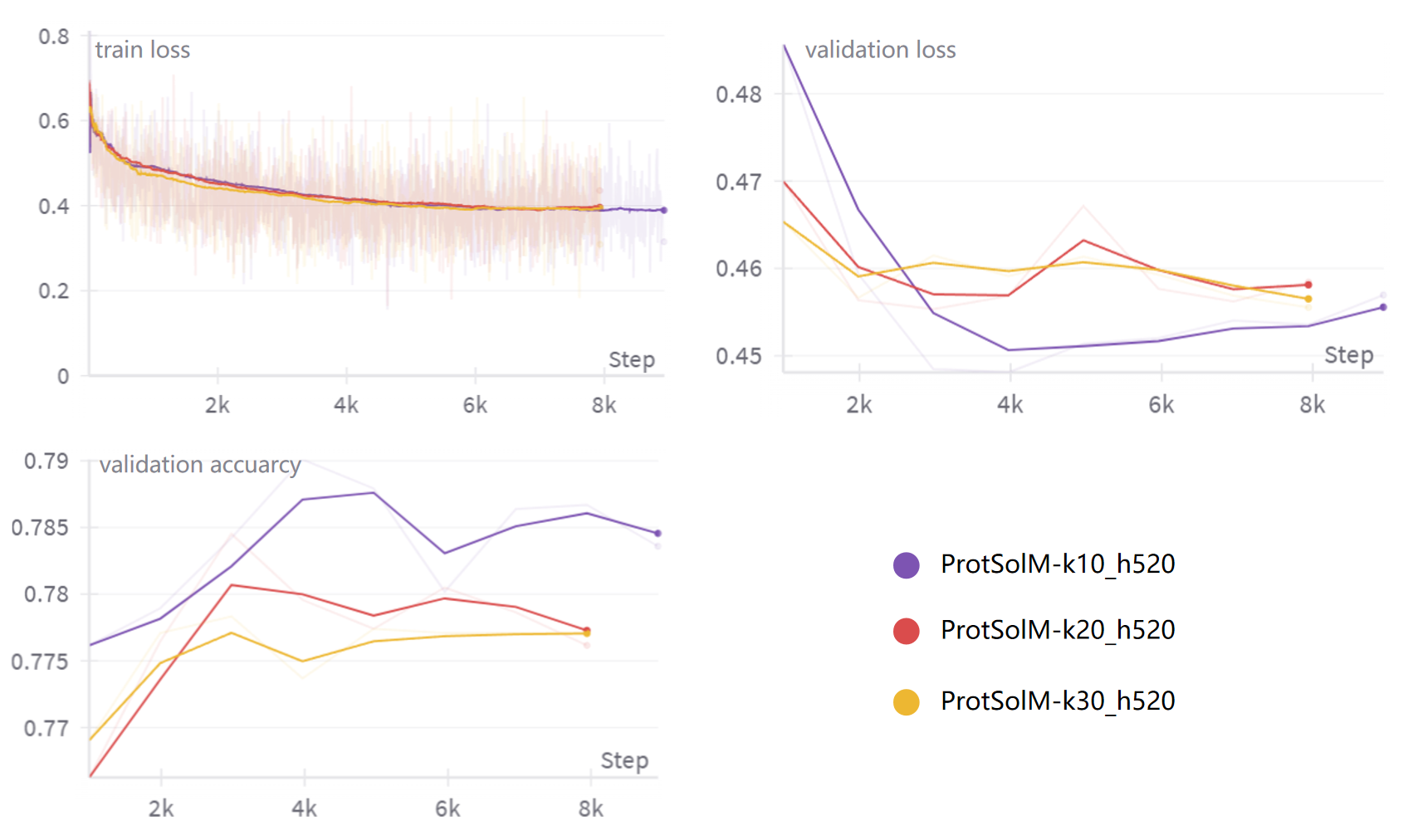}
    \caption{Learning curves of different \sol~models.}
    \label{fig:model_curve}
\end{figure}

From Fig.~\ref{fig:model_curve}, we can observe that on the validation dataset, as the number of neighbors increases when using KNN to build the graph from a protein structure, the validation performance decreases, but this is not observed on the test dataset.

\begin{figure}[h]
    \centering
    \includegraphics[width=0.9\linewidth]{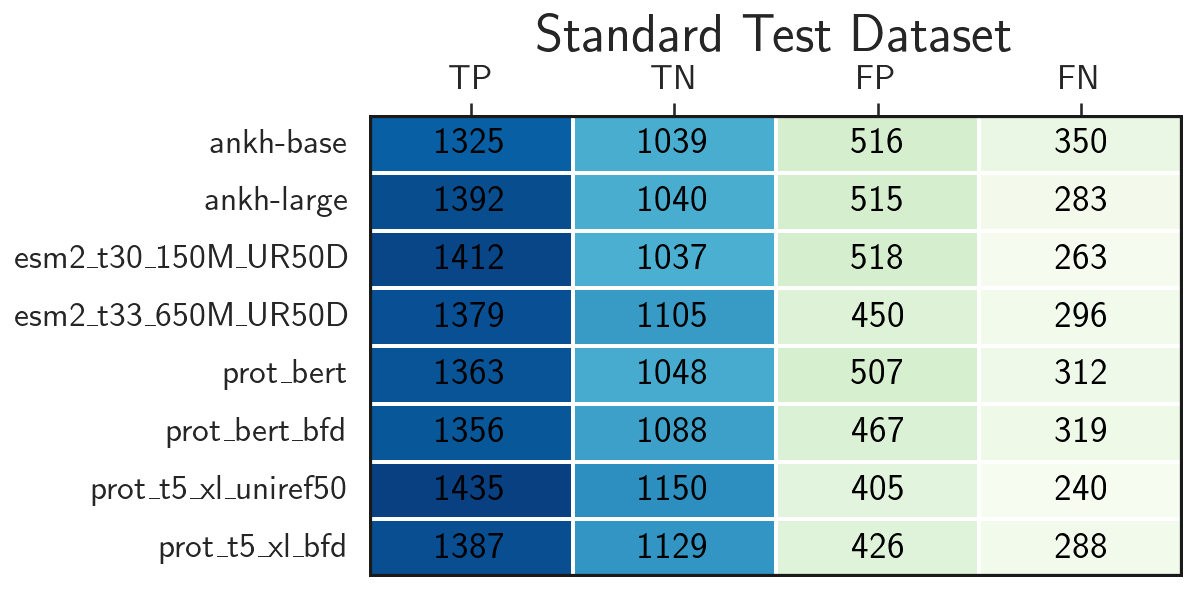}\\
    \vspace{2mm}
    \includegraphics[width=0.9\linewidth]{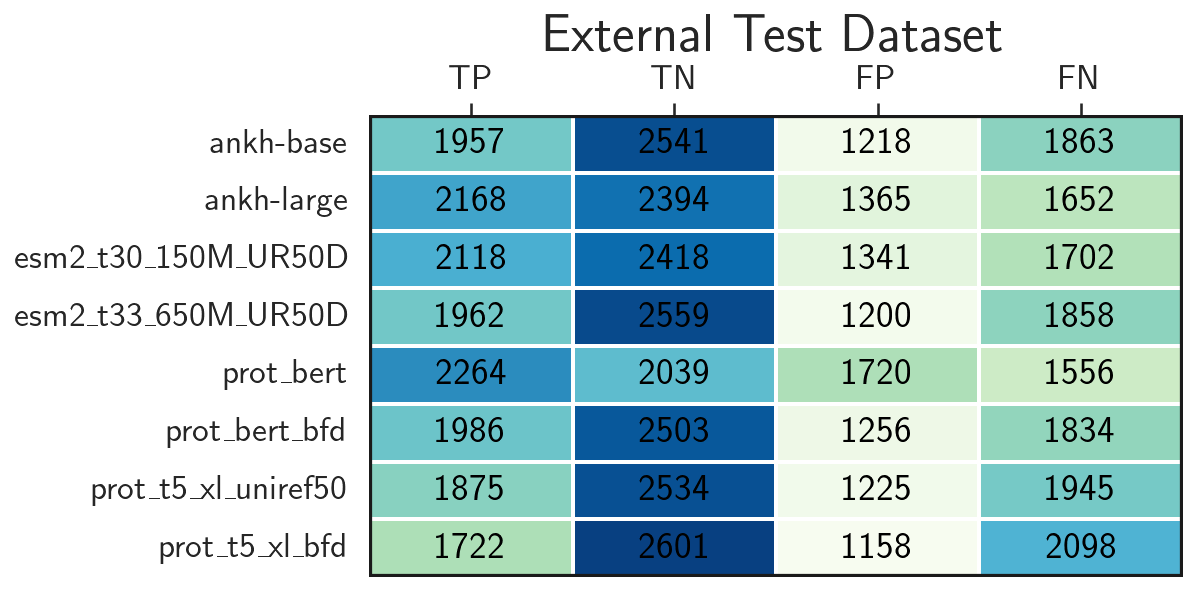}
    \caption{Confusion matrix of fine-tuned protein language models.}
    \label{fig:confusion_plm}
\end{figure}

In supplement to Fig.~\ref{fig:confusion_heatmap} in the main text, Fig.~\ref{fig:confusion_plm} reports the class-wide detailed prediction performance of fine-tuned protein language models.  

\begin{table}[!h]
\caption{Computation cost of different models on A800 GPU.}
\label{tab:time}
\begin{center}
\begin{tabular}{lcr}
\toprule
\textbf{Model} & \textbf{Version} & Time Cost  \\ 
\midrule
\multirow{2}{*}{ESM2}     & t30\_150M    & 19h 56m \\
                          & t33\_650M    & 36h 47m \\
\multirow{2}{*}{ProtBert} & uniref       & 24h 01m \\
                          & bfd          & 32h 48m  \\
\multirow{2}{*}{ProtT5}   & xl\_uniref50 & 124h 57m  \\
                          & xl\_bfd      & 124h 58m  \\
\multirow{2}{*}{Ankh}     & base         & 50h 14m  \\
                          & large        & 84h 43m \\
\multirow{3}{*}{\sol} & k10\_h512    & 26h 07m   \\
                          & k20\_h512    & 27h 28m  \\
                          & k30\_h512    & 32h 21m  \\ 
\bottomrule
\end{tabular}
\end{center}
\end{table}

Table~\ref{tab:time} lists the computational time required for fine-tuning a pre-trained model on \data-train. While the majority of models require 20-50 hours to complete the training procedure, our \sol~is one of the best-performing models that finished the fine-tuning process the fastest.

\section{Baseline Implements}
\label{app:source}
We list in Table~\ref{tab:BaselineAvailability} the open-sourced programs or web servers we used for training and testing the baseline methods. 

\begin{table}[!ht]
\caption{Baseline Availability}
\label{tab:BaselineAvailability}
\begin{center}
\resizebox{\linewidth}{!}{
\begin{tabular}{ll}
    \toprule
    \textbf{Method} & \textbf{Source}\\
    \midrule
     DeepSoluE \cite{DeepSoluE}& \url{http://lab.malab.cn/~wangchao/softs/DeepSoluE/}\\
     ccSOL omics \cite{ccSOL}& \url{http://s.tartaglialab.com/page/ccsol\_group}\\
     SoluProt \cite{Soluprot}& \url{https://loschmidt.chemi.muni.cz/soluprot/}\\
     SKADE \cite{SKADE}& \url{https://bitbucket.org/eddiewrc/skade/src/}\\
     Camsol \cite{CamSol}& \url{https://www-cohsoftware.ch.cam.ac.uk/index.php/}\\
     NetSolP \cite{NetSolP}& \url{https://services.healthtech.dtu.dk/services/NetSolP-1.0/}\\
     DSResSOL \cite{DSResSol}& \url{https://www.mdpi.com/article/10.3390/ijms222413555/s1}\\
     \midrule
     ESM2 \cite{esm2} & \url{https://huggingface.co/facebook/esm2_t33_650M_UR50D} \\
     ProtBert \cite{elnaggar2021prottrans} & \url{https://huggingface.co/Rostlab/prot_bert} \\
     ProtT5 \cite{elnaggar2021prottrans} & \url{https://huggingface.co/Rostlab/prot_t5_xl_uniref50} \\
     Ankh \cite{elnaggar2023ankh} & \url{https://huggingface.co/ElnaggarLab/ankh-base} \\
    \bottomrule
\end{tabular}
}
\end{center}
\end{table}

\end{document}